\documentclass[]{interact}

\usepackage{epstopdf}
\usepackage[caption=false]{subfig}

\usepackage[natbibapa,nodoi]{apacite}
\theoremstyle{plain}

\theoremstyle{definition}

\theoremstyle{remark}

\usepackage{multirow} 
\usepackage{xcolor} 

\begin{document}



\title{These Deals Won't Last! Longevity, Uniformity and Bias in Product Badge Assignment in E-Commerce Platforms}

\author{
\name{Archit Bansal\textsuperscript{a}, Kunal Banerjee\textsuperscript{b}
\thanks{This is an independent research done by Kunal Banerjee and not endorsed by Walmart in any manner.} 
and Abhijnan Chakraborty\textsuperscript{a}}
\affil{\textsuperscript{a}Indian Institute of Technology Delhi, India \\
\textsuperscript{b}Walmart Global Tech, Bangalore, India}
}

\maketitle

\begin{abstract}
Product badges are ubiquitous in e-commerce platforms, acting as effective psychological triggers to nudge customers to buy specific products, boosting revenues. However, to the best of our knowledge, there has been no attempt to systematically study these badges and their several idiosyncrasies -- we intend to close this gap in our current work.
Specifically, we try to answer questions such as: How long does a product retain a badge on a given platform? If a product is sold on different platforms, then does it receive similar badges?
How do the products that receive badges differ from those which do not, in terms of price, customer rating, etc. We collect longitudinal data from several e-commerce platforms over 45 days, and find that although most of the badges are short-lived, there are several permanent badge assignments and that too for badges meant to denote urgency or scarcity. Furthermore, it is unclear how the badge assignments are done, and we find evidence that highly-rated products are missing out on badges compared to lower quality ones. Our work calls for greater transparency in the badge assignment process to inform customers, as well as to reduce  dissatisfaction among the sellers dependent on the platforms for their revenues.
\end{abstract}


\section{Introduction}
\label{sec:intro}
\noindent 
Today, e-commerce marketplaces like Amazon, eBay, Rakuten and Alibaba have emerged as indispensable online platforms helping millions of customers with their purchase needs. At the same time, they provide livelihood to thousands of sellers and producers worldwide~\citep{smallSellerEcommerce}. Especially, during the pandemic induced restrictions on physical sales, these e-commerce marketplaces have become the lifeline for numerous small sellers~\citep{PTI2021Lockdown}. Given the sheer scale of the product space in these platforms,  algorithmic systems, such as search and recommendation systems, mediate the interactions between customers and sellers, deciding the customer experience and determing the fate of the sellers~\citep{sorokina2016amazon}. 

When a customer visits an e-commerce website and searches for something to buy, products matching the search query are returned. Interestingly, the returned product lists often contain products highlighted with certain badges, such as, `Best Seller', `Deal of the Day', `New Arrival', `Discount', etc. These badges evoke specific psychological reactions in a customer and nudge her to buy the corresponding products. For example, a `Best Seller' badge means that many other customers have already bought this popular product and hence, 
one can go forward and buy it. Whereas, the badge `Deal of the Day' creates a sense of urgency that the customer may miss out on a wonderful opportunity if that product is not bought immediately. 
In fact, product badges have been found to increase purchase conversion rates by as high as $55\%$~\citep{conversionBlog,yithBadge}. A specific use case of the cosmetics brand Ulta has been reported by~\citet{top3Blog}, where product badges have played a key role in boosting the company's overall revenue by $23\%$. 

Recognizing the importance, several Software-as-a-Service (SaaS) providers for smaller e-commerce businesses, including WooCommerce~\citep{wooCommerce}, BigCommerce~\citep{bigCommerce}, Crobox~\citep{crobox}, and PrefixBox~\citep{prefixBox}, put major emphasis on product badge assignment. For example, 
Crobox mentions that effective badge assignment may give an edge to smaller retailers to maintain competition with bigger players like Amazon~\citep{smallRetailerBadge}. These commercial tools allow creation of a number of badges, and may include the option of automatically assign badges to products~\citep{yithBadge}. 
The time duration for which a badge can be displayed on a platform can also be controlled by these tools.  

However, since the badges can be assigned to only a handful of products without risking the dilution of customer attention, it may deprive some products from getting the attention they truly deserve.
Specially, in some platforms, getting a badge can bring huge revenue opportunity for certain products.  
For example, the products with `Amazon's Choice' badges can be bought on Amazon with zero clicks! If a customer uses a voice-activated device, such as, Echo or Alexa, and says ``Alexa, order a pillow'', then the `Amazon's Choice' product would be the default choice for that keyword~\citep{zeroClick}. Interestingly, although some patterns have been recorded for which products have received the `Amazon's Choice' label, it is unclear to sellers that how Amazon's algorithm chooses which products should receive this badge~\citep{amazonChoiceUnclear}. It has been reported by~\citet{1pVs3p} that Amazon may provide preferential treatment to its own private label products over third-party sellers' products during recommendation; a lack of transparency in badge assignment may further fuel the dissatisfaction among the sellers.

Despite its 
ubiquitous presence and associated implications, only few research studies have focused on product badges. 
Adaji et al.~\citep{persuasive20,persuasive21} discussed different persuasive strategies employed in e-commerce platforms to boost sales, such as star ratings, votes, or reviews, but disregarded the role product badges play. A blogpost~\citep{badgeBlog} provided the first comprehensive categorization of different badges displayed at e-commerce websites, and our categorization scheme for badges borrows from this line of work. However, there is no cross-platform study looking at 
different aspects of product badges -- how long a product retains a particular badge, whether the platforms utilize badges for mis-selling, for example, a product always getting a badge {\it `$10\%$ off only for today'}, whether a product receives a badge uniformly across platforms, and so on. 

In this work, we bridge this gap by extensively collecting data from $12$ e-commerce platforms over $45$ days. Out of these $12$ platforms, $10$ cater to only niche domains, and rest $2$ are generic which sell a wide variety of products across multiple domains. 
All of our data is made publicly available\footnote{\url{https://github.com/architbansal28/Product-Badges}}  which, to the best of our knowledge, is the first comprehensive dataset that focuses on product badges, and includes details such as, name of the e-commerce platform, query executed, returned results ranked in order with brand names, badges assigned, prices, discounts, ratings, links and other useful relevant features.
Using this data, we attempt to answer the following three research questions:

\begin{itemize}
\item[] \textbf{RQ1. Longevity: } How long does a product retain a badge on a given platform? Does it vary depending on the platform, domain or the type of badges?
\item[] \textbf{RQ2. Uniformity:} If a product is sold on different platforms, does it receive similar badges?
\item[] \textbf{RQ3. Assignment bias:} How do the products that receive badges differ from those which do not? What is so special about these products with badges?
\end{itemize}

Our investigation reveals that though most of the badges are short lived, there are several products which retained a badge over the full data collection period. Most surprising are the badges denoting scarcity or urgency; one might expect them to be short lived, yet we find these assignments to be permanent. Same products seem to get different type of badges in different platforms; so there is no cross-platform consensus in badge assignments. We did not find any clear pattern in the product choices for badge assignment; in fact, products without badges seem to be higher rated than the products with badges. Overall, we take the first step towards exploring the badge landscape in this work, and we hope it can spawn future works looking at product badges in a holistic manner.

\section{Related Work}
\label{sec:relatedWork}
In this section, we briefly survey the prior works on e-commerce platforms and digital nudging.

\subsection{E-commerce platform designs}
E-commerce websites deploy multiple techniques to attract and retain customers for increasing their revenue, which include persuasive strategies targeting human psychology~\citep{influenceBook} or attractive interface design~\citep{vergo2002commerce}. Furthermore, customers' social networks are also utilized for boosting sales, in a form of e-commerce named as `Social Commerce'~\citep{xu2019think, 10.1145/3415185}. Having realized the potency of these strategies, many SaaS companies~\citep{bigCommerce, crobox} incorporate these in their services to small e-retailers that include search engine optimization, hosting, marketing, etc. 

Even within a platform, there are various means to steer a customer's attention to a specific product, which may include explicit identifiers, such as, stars, votes, reviews and badges, along with some implicit ones, such as, the rank of the product in the recommendation list~\citep{starVoteBadge}. However, unlike the rest of the explicit identifiers, badges are conferred to a chosen few products.
\citet{superhostAirbnb} looked into how the `SuperHost' badge in Airbnb helps improve sales at a higher charge. 
There are external companies which provide specialized badges to e-commerce websites, such as, Shopify~\citep{shopify} that provides payment related trust and security badges, and Yotpo~\citep{yotpo} that provides product reviews related badges.
~\citet{hallOfFameAmazon} reported how Amazon persuades its customers to write more reviews by giving the `Hall of Fame' badge. They further looked into various persuasive strategies that are employed in e-commerce~\citep{persuasive20,persuasive21}; however, they 
surprisingly did not cover product badges which are a common sight across all e-commerce platforms, and can be an effective tool 
to steer customers towards particular products. 

\subsection{Digital nudging}
\citet{thalernudge} introduced nudging as a tool to achieve societal goals by exploiting some mental shortcuts people take. Following this work, there have been many applications of nudge theory in several domains, including in digital platforms. For example,~\citet{10.1145/3415197} designed interface nudges to steer more donations to poorer schools in an educational charity platform.~\citet{10.1145/3479571} designed nudges to increase consumption of credible news on Twitter, which can help to curb misinformation. 
A comprehensive survey on digital nudging 
can be found by~\citet{surveyNudging}. According to their taxonomy, product badges should fall under the category `Attracting Attention', which, in turn, falls under the super-category `Increase salience of information'. However, the authors did not talk about product badges other than in a footnote, where they describe product badges as `visual highlighting tools'. 

\citet{persuasiveAd} tried to categorize different persuasive techniques in e-commerce. 
Unfortunately, product badges have been skipped in this work as well -- in fact, product badge does not belong to any of the categories mentioned by the authors, although `visual layout' category comes close.
Another way to nudge customers to buy products borrows ideas from gamification~\citep{gamification}, such as, through loyalty points, coax customers to enter contests, etc. Some researchers have also criticised the  interface design choices of e-commerce platforms to push users into making unintended purchases as {\it dark patterns}~\citep{10.1145/3359183}. 

\vspace{1mm}
\noindent Overall, we see that psychologically targeting customers and subtly persuading them towards purchasing specific products is prevalent in e-commerce; however, product badges seem to be a comparatively new phenomenon and not well explored in the literature. In this work, we attempt to bridge this gap.
\section{Dataset Gathered}
\label{sec:data}
\noindent 
As mentioned in the introduction, in this work, we attempt to explore the badge assignments 
across different e-commerce platforms. For this purpose, we 
looked into platforms that serve a niche domain vis-a-vis generic platforms that sell a variety of products across domains, including the niche ones. 
We are also interested to know whether the same products 
sold on different platforms get similar badges. 
Therefore, we restricted ourselves to websites that target a specific country; otherwise 
the products (brands) being sold might be quite different. 
In this study, we have chosen websites that are operational in India.

We concentrate our study on four niche domains: (i) baby products, (ii) cosmetics, (iii) fashion, and (iv) home decor. These domains were chosen based on two factors. 
Firstly, we tried to make the niche domains as diverse as possible. Though there may be small overlaps,  e.g., websites related to baby products and fashion may both include garments for babies; we carefully selected the queries to collect data from different websites 
so that the 
returned product lists are different across domains. 
Secondly, the selected domains should have websites that are popular in India. 

For each domain, we picked two or three niche websites, totaling ten websites across four domains, as listed in Table~\ref{tab:dataset1}. 
As generic websites, we chose \textit{Amazon} and \textit{Snapdeal}, which sell products across all these domains.\footnote{It is worth noting that for platforms with a global presence, we have considered their India specific website, e.g., {\tt Amazon.in}.} 
Next, we selected ten different queries 
for each of these four domains so that the corresponding e-commerce websites (niche and generic) 
returned a considerable number of products as results. 
These queries are also reported in Table~\ref{tab:dataset1}.

\begin{table*}[t!]
\centering
\caption{\bf Data collection: e-commerce websites from different domains and the queries we used to collect the corresponding product listing.}
\vspace{1mm}
\resizebox{.98\linewidth}{!}
{
\begin{tabular}{|l|l|p{2.7cm}|p{8cm}|}
\hline
{\bf Domain} & {\bf Niche Websites} & {\bf No. of unique products} & {\bf Search Queries}\\
\hline
\multirow{3}{*}{Baby Products} & FirstCry ({\tt firstcry.com}) & 4145
& Baby diaper, Baby toys, Baby boy clothes, \\ 
& Hopscotch ({\tt hopscotch.in}) & 1196 & Stroller, Baby towel, Baby shampoo, Baby girl\\ 
& My Baby Babbles ({\tt mybabybabbles.com}) & 972 & dresses, Onesies, Baby footwear, Baby food \\ \hline

\multirow{3}{*}{Cosmetics} & e.l.f. Cosmetics ({\tt elfcosmetics.com}) & 226
& Eyeliner, Lipstick, Blush, Cleanser, Moisturizer,\\ 
& Nykaa ({\tt nykaa.com}) & 1413 & Face primer, Makeup removers, Face brush, Lip\\
& & & balm, Mascara \\\hline

\multirow{4}{*}{Fashion} & Bewakoof ({\tt bewakoof.com}) & 1275
& Women dresses, Women kurtis, Women jackets, \\
& LimeRoad ({\tt limeroad.com}) & 6536 & Women jeans, Women shirts, Men casual shirts,\\
& Max Fashion ({\tt maxfashion.in}) & 1024 & Men trousers, Men polo t-shirts, Men socks, Men\\ 
& & & track pants\\\hline

\multirow{3}{*}{Home Decor} & Pepperfry ({\tt pepperfry.com}) & 2319
& Bed, Dining set, Mattress, Bedsheet, Table \\
& Urban Ladder ({\tt urbanladder.com}) & 3498 & lamp, Sofa set, Study table, Wall mirror, Chair, \\
& & & Artificial plants \\
\hline \hline

{\bf Domain}  & {\bf Generic Websites} & {\bf No. of unique products} & {\bf Search Queries} \\ \hline
\multirow{2}{*}{All four} & Amazon ({\tt amazon.in}) & 23784
& \multirow{2}{*}{All queries mentioned above} \\
& Snapdeal ({\tt snapdeal.com}) & 16501 & \\
\hline
\end{tabular}}
\label{tab:dataset1}
\end{table*}

\if(0)
\begin{table*}[t!]
\centering
\caption{\bf Number of products from different e-commerce websites present in our dataset.}
\resizebox{.95\linewidth}{!}{
\begin{tabular}{|l|l|p{6cm}|r|}
\hline
{\bf E-commerce website} & {\bf No. of unique products} & {\bf No. of products which got at least one badge during the 45 day period} & {\bf \%}\\
\hline
Amazon ({\tt amazon.in})                  & 23784 & 6124 & 25.7 \\
Bewakoof ({\tt bewakoof.com})             & 1275  & 781  & 61.2 \\
e.l.f. Cosmetics ({\tt elfcosmetics.com}) & 226   & 66   & 29.2 \\
FirstCry ({\tt firstcry.com})             & 4145  & 2028 & 48.9 \\
Hopscotch ({\tt hopscotch.in})            & 1196  & 607  & 50.7 \\
LimeRoad ({\tt limeroad.com})             & 6536  & 4713 & 72.1 \\
Max Fashion ({\tt maxfashion.in})         & 1024  & 886  & 86.5 \\
My Baby Babbles ({\tt mybabybabbles.com}) & 972   & 119  & 12.2 \\
Nykaa ({\tt nykaa.com})                   & 1413  & 1312 & 92.8 \\
Pepperfry ({\tt pepperfry.com})           & 2319  & 661  & 28.5 \\
Snapdeal ({\tt snapdeal.com})             & 16501 & 8220 & 49.8 \\
Urban Ladder ({\tt urbanladder.com})      & 3498  & 1499 & 42.8 \\
\hline
\end{tabular}}
\label{tab:dataset2}
\end{table*}
\fi

To automatically collect the product listing against each query, we used the Selenium WebDriver ({\tt selenium.dev}) to automate the process of firing the search query to these websites and 
retrieve the results returned. We considered the results up to a maximum of 100 products, and also stored other metadata associated with the products, 
such as product name, price, discount, average customer rating, number of ratings, and most importantly, the badge assigned to a product (if available). 
We call one instance of the data collected across all websites as a \textit{snapshot}. 
Our dataset contains data collected over 45 days during June--August 2021, with 2 snapshots per day 
at a gap of 12 hours, i.e., altogether there are total $90$ snapshots. 
Overall, the data consists of $62,889$ unique products across platforms, out of which $27,016$ ($42.9\%$) 
products got some badge in at least one of the snapshots. 

\section{Categorization of Badges}
\label{sec:category}

\begin{table*}[t!]
\centering
\caption{\bf Badges assigned at different e-commerce websites and their respective categories. Here `X' and `Y' represent different numbers used in the actual badges -- for example, `Only 5 Left in Stock', `Buy 3 Get 2',  etc.}
\vspace{1mm}
\resizebox{0.98\linewidth}{!}
{
\begin{tabular}{|l|l|l|}
\hline
{\bf Website} & {\bf Badge} & {\bf Category}\\
\hline
\multirow{5}{*}{Amazon} & Best Seller, Kids Gift Ideas & Social Proof\\
                        & Only X Left in Stock        & Scarcity\\
                        & Deal of the Day, Limited Time Deal, Deal is X\% Claimed, Prime Day Deal & Urgency\\
                        & Amazon’s Choice             & Endorsement\\
                        & Prime Day Launch            & Exclusivity\\
\hline
\multirow{3}{*}{Bewakoof} & Few Left, Last Sizes Left - Special Price & Scarcity\\
                          & Flash Sale & Urgency\\
                          & Buy X Get Y, Buy X for Y, Color of the Month & Promotional\\
\hline
\multirow{3}{*}{e.l.f. Cosmetics} & Best Sellers, Trending, Glam Award & Social Proof\\
                              & New        & Recency\\
                              & Holy Grail & Endorsement\\
\hline
\multirow{4}{*}{FirstCry} & Bestseller, \#1 Mom's Pick & Social Proof\\
                          & X Left & Scarcity\\
                          & New!   & Recency\\
                          & Super Saver & Promotional\\
\hline
Hopscotch & X Left & Scarcity\\
\hline
\multirow{4}{*}{LimeRoad} & Flash Sale, Freshness Unplugged Season Sale, Brand Day Sale & Urgency\\
                          & New       & Recency\\
                          & Offer: Buy X Get Y Free, Offer: Freebie, Best Price & Promotional\\
                          & Exclusive & Exclusivity\\
\hline
\multirow{2}{*}{Max Fashion} & New         & Recency\\
                            & Buy X Get Y, Buy X at Y, Buy X at Y\% Off, Flat X\% Off & Promotional\\
\hline
My Baby Babbles & Sale & Urgency\\
\hline
\multirow{5}{*}{Nykaa} & Bestseller & Social Proof\\
                       & Sale       & Urgency\\
                       & New        & Recency\\
                       & Offer      & Promotional\\
                       & Featured   & Endorsement\\
\hline
\multirow{4}{*}{Pepperfry} & Best Seller          & Social Proof\\
                           & Clearance Sale!     & Urgency\\
                           & New                 & Recency\\
                           & 30/100 Nights Trial & Promotional\\
\hline
\multirow{3}{*}{Snapdeal} & Trending, Top Seller, X Orders in Last Y Days, X People Just Ordered, X\% Positive Feedback & Social Proof\\
                          & X Left!  & Scarcity\\
                          & Featured & Endorsement\\
\hline
\multirow{3}{*}{Urban Ladder} & Best Seller & Social Proof\\
                              & Only X Left & Scarcity\\
                              & New Arrival & Recency \\
\hline
\end{tabular}}
\label{tab:badge}
\end{table*}

As discussed in the last section, we explore a total of $12$ e-commerce websites, and all of these assign multiple types of badges to different products, where these badge names are often unique to the site. To perform a cross-platform study, we group different badges into coherent categories. 
Badges are designed to trigger specific psychological reactions from the customers,
so that they take a mental note of the products with the badges and 
get swayed towards buying them. Following the lines of~\citet{influenceBook} and~\citet{badgeBlog}, we categorize the badges into the following categories, based on the psychological triggers they fire. 

\begin{itemize}
\item[] \textbf{Social Proof:} These badges indicate that the product is popular among customers, e.g., `Best Seller', `Trending'.
\item[] \textbf{Scarcity:} This type of badges communicate that the product is only available in a small  quantity, e.g., `Few Left', `Only 5 Left in Stock'.
\item[] \textbf{Urgency:} These badges convey that one needs to buy the product quickly, else she will miss a great opportunity, e.g., `Deal of the Day', `Limited Time Deal'.
\item[] \textbf{Recency:} This type of badges indicate that the product has been introduced recently, e.g., `New Arrival', `New'. 
\item[] \textbf{Promotional:} This type of badges indicate that by buying the product now, a customer will receive a special offer or discounted price, e.g., `Buy 1 Get 1', `Super Saver'. 
\item[] \textbf{Endorsement:} These badges carry the endorsement from the platform or some expert, e.g., `Amazon's Choice', `Featured'.
\item[] \textbf{Exclusivity:} This type of badges indicate that the product is available exclusively on a platform or for a chosen set of customers, e.g., `Exclusive', `Prime Day Launch'.
\end{itemize}

It is worth noting that a type of badge may invoke a mix of emotions. 
For example, 
a \textit{Scarcity} badge 
also indicates that the product should be bought urgently and hence can also be put in \textit{Urgency} category. Furthermore, 
a \textit{Scarcity} badge also tacitly implies that the product has been 
selling fast, and thus serves as a \textit{Social Proof}. 
Similarly, a badge like `10 People Just Ordered' can belong to both \textit{Social Proof} and \textit{Recency} category. To reduce the chance of misinterpretation, three annotators were asked to independently assign categories to the badges we encountered in our dataset, by examining their nomenclature, and we choose the category of a badge based on the majority opinion. Table~\ref{tab:badge} lists all badges present in different websites and their respective category information. 
Note that we did not include the `Sponsored' badge in our study because 
it is assigned in exchange for a fee, and thus does not shed any insight on the badge assignment policies. 
For similar reasons, we also did not consider `Out of Stock' and `Sold Out' badges in our study.


\section{RQ1. Longevity: How long do the deals last?}
After categorizing the badges, we turn our attention to the {\it stability of badge assignments} at different e-commerce platforms. Knowing how long a product holds a particular badge (termed as `Longevity') is important for multiple reasons: (i) if a \textit{Scarcity} or \textit{Urgency} category badge is given to a product for long, it can indicate potentially unfair practice by the platform; retaining the same badge to particular products (ii) may lose its value for the repeat customers, and (iii) may deny opportunities to other products.

\subsection{Measuring longevity}
We compute longevity using two metrics: Occurrence and Consistency. 
Occurrence of a badge $i$ for a product $j$ ($\mathcal{O}_{i,j}$) is defined as 
$$
\mathcal{O}_{i,j} = \frac{\textnormal{\# snapshots with } j \textnormal{ having badge } i}{\textnormal{total number of snapshots}}.
$$
For example, if Product A has the badge X for $60$ snapshots out of $90$, then 
$\mathcal{O}_{X,A}$ is $\frac{60}{90} = 0.67$. Average occurrence of a badge $i$ ($\hat{\mathcal{O}}_{i}$) is calculated over all products having this badge:
$$
\hat{\mathcal{O}}_{i} = \frac{\sum_j \mathcal{O}_{i,j}}{\sum_j \mathcal{I}{(\mathcal{O}_{i,j} > 0})}, 
$$
where $\mathcal{I}$ is the indicator function.

Additionally, we use another metric {\it Consistency} ($\mathcal{C}_{i,j}$) as a measure of how long a product $j$ retains badge $i$ continuously. This is related to {\it Occurrence}, but here the focus is on the contiguous time stretch. We measure $\mathcal{C}_{i,j}$ as 
$$
\mathcal{C}_{i,j} = \frac{max (\textnormal{\# contiguous snapshots with } j \textnormal{ having badge } i)}{\textnormal{total number of snapshots}}.
$$
For example, if a platform assigns badge $X$ to product $A$ in snapshot 1, 2 and 3, but not in 4, $\mathcal{C}_{X,A}$ would be $\frac{3}{4} = 0.75$.

\begin{figure*}[!t]
\centering
\includegraphics[width=0.99\linewidth]{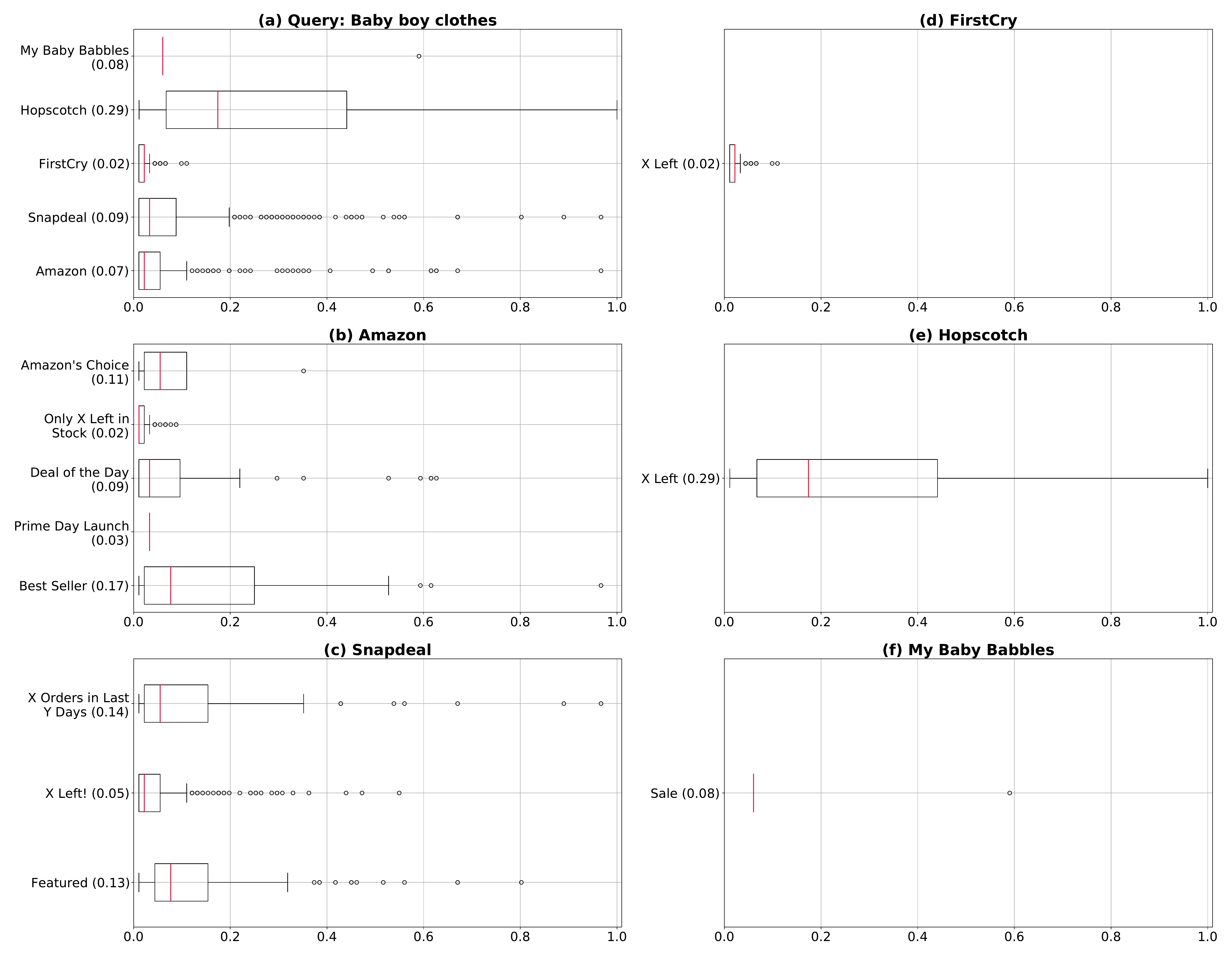}
\caption{\bf Box plots of Occurrence ($\mathcal{O}$) values for the query `baby boy clothes' under \textit{Baby Products}.
Interestingly, every niche platform provided a single badge: either \textit{X Left (Scarcity)} or \textit{Sale (Urgency)}; whereas, the generic platforms presented a range of badge categories: \textit{Social Proof, Scarcity, Urgency, Endorsement, Exclusivity}.}
\label{fig:baby_clothes}
\end{figure*}

\begin{figure*}[!t]
\centering
\includegraphics[width=0.99\linewidth]{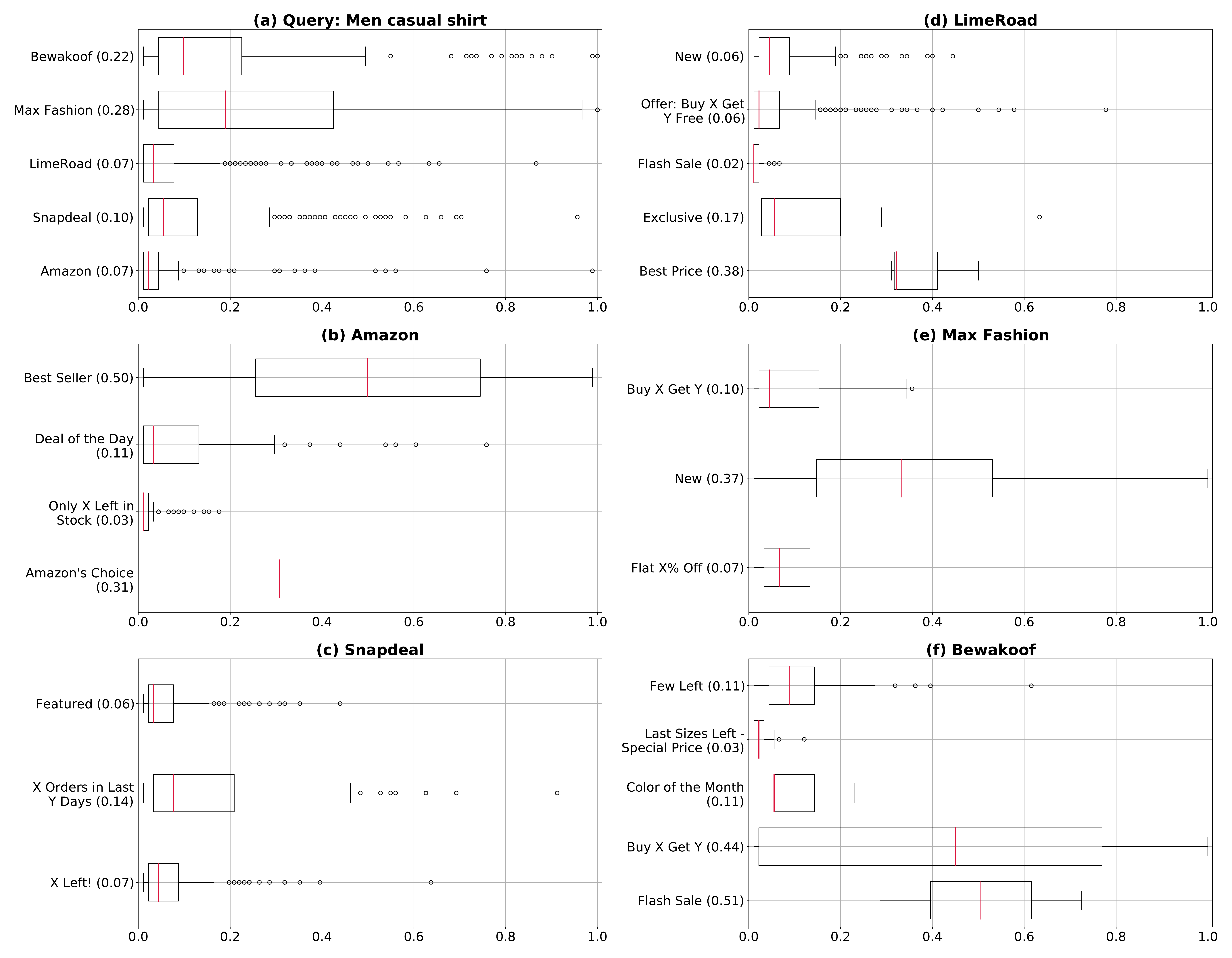}
\caption{\bf Box plots of Occurrence ($\mathcal{O}$) values for the query `men casual shirt' under \textit{Fashion}.
For this query, both generic and niche platforms provided similar number of badges over a similar spectrum of badge categories.}
\label{fig:men_shirt}
\end{figure*}

\subsection{Longevity of product badges across platforms}
Box plots in Figure~\ref{fig:baby_clothes} and Figure~\ref{fig:men_shirt} present the range of $\mathcal{O}_{i,j}$ values across products and badges in different platforms 
for two queries, with $\hat{\mathcal{O}}_{i}$ values shown in the parenthesis. 
While Figure~\ref{fig:baby_clothes}(a) and Figure~\ref{fig:men_shirt}(a) show the $\mathcal{O}_{i,j}$ values across all badges in a platform for a query (values in parenthesis denoting average  $\hat{\mathcal{O}}_{i}$), all other sub-figures show $\mathcal{O}_{i,j}$ values for individual badges in respective platforms. 
It is interesting to note that for the query `baby boy clothes', every niche platform provided a single badge: either \textit{X Left (Scarcity)} or \textit{Sale (Urgency)}; whereas, the generic platforms 
presented a range of badge categories: \textit{Social Proof, Scarcity, Urgency, Endorsement, Exclusivity}.
On the other hand, for `men casual shirt' query, both generic and niche platforms provided similar number of badges over a similar spectrum of badge categories.

Furthermore, in Figure~\ref{fig:baby_clothes}, Hopscotch is seen to give away badges with the highest longevity, with some items having the maximum possible longevity $1$.
The badges in FirstCry have the lowest longevity, on an average, whereas, My Baby Babbles has given the least amount of badges.
Snapdeal, although it gave less type of badges compared to Amazon, had badges with higher longevity than the other generic platform.
In Figure~\ref{fig:men_shirt}, we see that Max Fashion, on an average, has badges with highest longevity; in some cases, the longevity is $1$ for Max Fashion and Bewakoof.
Diving in, for Max Fashion, \textit{New (Recency)} badges, and for Bewakoof, \textit{Buy X Get Y (Promotional)} badges are the longest lasting.
Amazon, on the other end, gave away badges with least average longevity.
However, the \textit{Best Seller (Social Proof)} badges in Amazon have high longevity, including $1$ in some cases; in contrast, \textit{Deal of the Day (Urgency)} and \textit{Only X Left in Stock (Scarcity)} badges have low longevity, thereby bringing down the average longevity for this platform.

After exploring the longevity of product badges at a query level, 
we aggregate the observations across all queries for a platform to identify macro trends.
Specifically, 
we first consider all products with $\mathcal{O}_{i,j} > 0$; 
next we consider only those products having $\mathcal{O}_{i,j} > 0.1$, 
and so on. Table~\ref{tab:longevity_website} shows these products across different websites under different domains. 
According to our data, of all the badges assigned, only $36.7\%$ of the products retained their badge for more than $10\%$ of the snapshots, indicating that the longevity of badges for the majority of products is rather small and there is a large churn in the assigned badges. Only $9.4\%$ of the products had their badges for more than $50\%$ snapshots.
When we look into the individual domains, it is hardest to retain a badge in \textit{Fashion}, while in \textit{Cosmetics}, the badges are retained fairly long. 
Of all the websites, e.l.f. Cosmetics gave the least number of badges -- this is probably because it sells products of only its own brand and consequently, the number of products sold here is also less.
On the other hand, LimeRoad gave the maximum number of badges.
Between the generic platforms, 
collectively Snapdeal assigned more badges than Amazon.

\begin{table*}[t]
\centering
\caption{\bf Longevity of badge assignments across different websites and domains. The values in parenthesis represent the average Occurrence and average Consistency respectively.
Among the individual domains, \textit{Fashion} has minimum badge retention while \textit{Cosmetics} has maximum.}
\setlength{\tabcolsep}{1pt}
\resizebox{0.99\linewidth}{!}
{
\begin{tabular}{|r|| r|r|r|r|| r|r|r|r|| r|r|r|r|| r|r|r|r|| r|r|r|r|}
\hline
$\mathcal{O}_{i,j}$ &
$>0$ & $>0.1$ & $>0.2$ & $>0.5$ &
$>0$ & $>0.1$ & $>0.2$ & $>0.5$ &
$>0$ & $>0.1$ & $>0.2$ & $>0.5$ &
$>0$ & $>0.1$ & $>0.2$ & $>0.5$ &
$>0$ & $>0.1$ & $>0.2$ & $>0.5$ \\
\hline
\hline
\textbf{Baby Products} & 
\multicolumn{4}{c||}{Amazon (0.12, 0.07)} & 
\multicolumn{4}{c||}{Snapdeal (0.12, 0.06)} & 
\multicolumn{4}{c||}{FirstCry (0.12, 0.09)} & 
\multicolumn{4}{c||}{Hopscotch (0.27, 0.21)} & 
\multicolumn{4}{c|}{My Baby Babbles} \\
& \multicolumn{4}{c||}{}&\multicolumn{4}{c||}{}&\multicolumn{4}{c||}{}&\multicolumn{4}{c||}{}& \multicolumn{4}{c|}{(0.51, 0.51)} \\
\hline
\#products &
1628 & 464 & 263 & 119 &
2599 & 725 & 447 & 155 &
2025 & 421 & 307 & 158 &
 599 & 362 & 247 & 106 &
 118 &  77 &  70 &  62 \\
\#badges &
6 & 5 & 5 & 4 &
6 & 4 & 4 & 4 &
5 & 5 & 5 & 5 &
1 & 1 & 1 & 1 &
1 & 1 & 1 & 1 \\
\hline
\hline
\textbf{Cosmetics} & 
\multicolumn{4}{c||}{Amazon (0.12, 0.07)} & 
\multicolumn{4}{c||}{Snapdeal (0.18, 0.09)} & 
\multicolumn{4}{c||}{Nykaa (0.32, 0.15)} & 
\multicolumn{4}{c||}{e.l.f. Cosmetics (0.50, 0.50)} & 
\multicolumn{4}{c|}{} \\
\hline
\#products &
1206 & 330 & 190 &  74 &
1005 & 453 & 314 & 104 &
1311 & 873 & 705 & 326 &
  66 &  42 &  38 &  27 &
\multicolumn{4}{c|}{} \\
\#badges &
5 & 5 & 5 & 4 &
5 & 4 & 4 & 4 &
5 & 4 & 4 & 4 &
5 & 5 & 5 & 5 &
\multicolumn{4}{c|}{} \\
\hline
\hline
\textbf{Fashion} & 
\multicolumn{4}{c||}{Amazon (0.06, 0.03)} & 
\multicolumn{4}{c||}{Snapdeal (0.11, 0.05)} & 
\multicolumn{4}{c||}{LimeRoad (0.09, 0.03)} & 
\multicolumn{4}{c||}{Max Fashion (0.33, 0.22)} & 
\multicolumn{4}{c|}{Bewakoof (0.33, 0.23)}\\
\hline
\#products &
1643 &  224 &  96 &  31 &
3444 &  956 & 538 & 170 &
4698 & 1206 & 569 & 127 & 
 862 &  587 & 422 & 244 &
 781 &  494 & 372 & 220 \\
\#badges &
5 & 5 & 4 & 3 &
5 & 4 & 4 & 4 &
5 & 5 & 4 & 4 &
3 & 3 & 2 & 2 &
5 & 5 & 5 & 4 \\
\hline
\hline
\textbf{Home Decor} & 
\multicolumn{4}{c||}{Amazon (0.07, 0.04)} & 
\multicolumn{4}{c||}{Snapdeal (0.21, 0.11)} & 
\multicolumn{4}{c||}{Urban Ladder (0.12, 0.06)} & \multicolumn{4}{c||}{Pepperfry (0.37, 0.30)} & 
\multicolumn{4}{c|}{} \\
\hline
\#products &
1599 & 311 & 140 &  36 &
1151 & 556 & 412 & 170 &
1499 & 558 & 306 &  39 &
 636 & 491 & 387 & 176 &
\multicolumn{4}{c|}{} \\
\#badges &
6 & 5 & 5 & 4 &
5 & 3 & 3 & 3 &
3 & 3 & 3 & 3 &
4 & 4 & 4 & 3 &
\multicolumn{4}{c|}{} \\
\hline
\end{tabular}}
\label{tab:longevity_website}
\end{table*}

\subsection{Longevity of different badge categories}
We now look into the longevity of the product badges based on the categories that these belong to.
Table~\ref{tab:longevity_badge} shows an overview of the same.
We see that most of the products with badges belong to \textit{Scarcity} -- 39.4\% of all such products, while \textit{Exclusivity} badges are the most rare -- only 0.4\%.
The longevity of the \textit{Scarcity} badges are the least with only 27.6\% products having a lifespan of more than 10\% snapshots.
We expected the \textit{Urgency} badges to follow a similar trajectory; however, to our surprise, 45.6\% of these badges are retained for more than 10\% snapshots, and 15.4\% are retained for more than 50\% snapshots, which is higher than \textit{Scarcity, Recency, Promotional, Endorsement} and \textit{Exclusivity} badges.

When we looked into the badge assignments across the domains, we found that except for \textit{Scarcity} and \textit{Urgency} categories, all the other types of badges had the maximum presence in the \textit{Fashion} domain.
In fact, 69.1\% of the products with \textit{Recency} badges and 75.7\% of the products with \textit{Promotional} badges are from the \textit{Fashion} domain.
The reason might be that compared to the other domains, the customers of the \textit{Fashion} industry are much more interested in keeping up with the latest trends (which explains \textit{Recency}), and the profit margins are much higher in apparels which allows the brands to run several \textit{Promotional} offers.
Interestingly, domain-wise, 42.5\% of the products that received a badge belong to \textit{Fashion}, and the second highest being \textit{Baby Products} with 24.3\% of the products; however, at $>$50\% level, \textit{Fashion} has the lowest badge retention rate of 8.5\%.

\begin{table*}[t]
\centering
\caption{\bf Longevity of badge assignments across different badge categories.
Surprisingly, \textit{Urgency} badges are retained for a long period.
Domain-wise, typically, \textit{Fashion} receives the maximum number of badges -- the longevity of these badges, however, are small.}
\setlength{\tabcolsep}{1pt}
\resizebox{0.99\linewidth}{!}
{
\begin{tabular}{|l|r|| r|r|r|r|| r|r|r|r|| r|r|r|r|| r|r|r|r|}
\hline
 & $\mathcal{O}_{i,j}$ &
$>0$ & $>0.1$ & $>0.2$ & $>0.5$ &
$>0$ & $>0.1$ & $>0.2$ & $>0.5$ &
$>0$ & $>0.1$ & $>0.2$ & $>0.5$ &
$>0$ & $>0.1$ & $>0.2$ & $>0.5$ \\
\cline{2-18}
Category & &
\multicolumn{4}{c||}{\textbf{Baby Products}} &
\multicolumn{4}{c||}{\textbf{Cosmetics}} &
\multicolumn{4}{c||}{\textbf{Fashion}} &
\multicolumn{4}{c|}{\textbf{Home Decor}} \\
\hline
\hline
\multirow{3}{*}{Social Proof} &
avg $\mathcal{O}$, avg $\mathcal{C}$ & 
\multicolumn{4}{c||}{(0.30, 0.19)} &
\multicolumn{4}{c||}{(0.40, 0.26)} &
\multicolumn{4}{c||}{(0.22, 0.10)} &
\multicolumn{4}{c|}{(0.32, 0.22)} \\ 
\cline{3-18}
& \#products & 
660 & 402 & 307 & 165 &
271 & 198 & 171 &  97 &
993 & 534 & 351 & 143 &
785 & 511 & 378 & 194 \\
& \#badges &
8 & 5 & 5 & 5 &
8 & 7 & 7 & 7 &
4 & 3 & 3 & 3 &
7 & 4 & 4 & 4 \\
\hline
\hline
\multirow{3}{*}{Scarcity} &
avg $\mathcal{O}$, avg $\mathcal{C}$ & 
\multicolumn{4}{c||}{(0.11, 0.07)} &
\multicolumn{4}{c||}{(0.13, 0.06)} &
\multicolumn{4}{c||}{(0.11, 0.06)} &
\multicolumn{4}{c|}{(0.13, 0.06)} \\
\cline{3-18}
& \#products &
5214 & 1273 & 799 & 328 &
1222 &  372 & 226 &  76 &
3193 &  876 & 509 & 187 &
2755 &  903 & 582 & 174 \\
& \#badges &
4 & 4 & 4 & 4 &
2 & 2 & 2 & 2 &
4 & 4 & 4 & 4 &
3 & 3 & 3 & 3 \\
\hline
\hline
\multirow{3}{*}{Urgency} &
avg $\mathcal{O}$, avg $\mathcal{C}$ & 
\multicolumn{4}{c||}{(0.21, 0.14)} &
\multicolumn{4}{c||}{(0.31, 0.15)} &
\multicolumn{4}{c||}{(0.11, 0.05)} &
\multicolumn{4}{c|}{(0.11, 0.06)} \\
\cline{3-18}
& \#products &
 974 &  437 & 286 & 162 &
1587 & 1011 & 803 & 376 &
1381 &  445 & 238 &  66 &
 748 &  246 & 118 &  35 \\
& \#badges &
2 & 2 & 2 & 2 &
2 & 1 & 1 & 1 &
3 & 3 & 2 & 2 &
2 & 2 & 2 & 1 \\
\hline
\hline
\multirow{3}{*}{Recency} &
avg $\mathcal{O}$, avg $\mathcal{C}$ & 
\multicolumn{4}{c||}{(0.21, 0.18)} &
\multicolumn{4}{c||}{(0.32, 0.28)} &
\multicolumn{4}{c||}{(0.16, 0.09)} &
\multicolumn{4}{c|}{(0.14, 0.10)} \\
\cline{3-18}
& \#products & 
 181 &  113 &  73 &  14 &
  94 &   53 &  43 &  21 &
2557 & 1007 & 575 & 254 &
 867 &  379 & 218 &  28 \\
& \#badges &
1 & 1 & 1 & 1 &
2 & 2 & 2 & 2 &
2 & 2 & 2 & 2 &
2 & 2 & 2 & 2 \\
\hline
\hline
\multirow{3}{*}{Promotional} &
avg $\mathcal{O}$, avg $\mathcal{C}$ & 
\multicolumn{4}{c||}{(0.40, 0.38)} &
\multicolumn{4}{c||}{(0.33, 0.14)} &
\multicolumn{4}{c||}{(0.17, 0.08)} &
\multicolumn{4}{c|}{(0.41, 0.38)} \\
\cline{3-18}
& \#products &
  41 &   16 &  16 &  16 &
1153 &  812 & 668 & 297 &
3926 & 1499 & 948 & 426 &
  68 &   58 &  57 &  11 \\
& \#badges &
1 & 1 & 1 & 1 &
1 & 1 & 1 & 1 &
6 & 5 & 5 & 4 &
1 & 1 & 1 & 1 \\
\hline
\hline
\multirow{3}{*}{Endorsement} &
avg $\mathcal{O}$, avg $\mathcal{C}$ & 
\multicolumn{4}{c||}{(0.22, 0.12)} &
\multicolumn{4}{c||}{(0.30, 0.16)} &
\multicolumn{4}{c||}{(0.10, 0.05)} &
\multicolumn{4}{c|}{(0.22, 0.13)} \\
\cline{3-18}
& \#products &
 552 & 272 & 195 &  76 &
 657 & 383 & 301 & 162 &
1248 & 310 & 169 &  60 &
 203 & 108 &  82 &  38 \\
& \#badges &
2 & 2 & 2 & 2 &
4 & 4 & 4 & 4 &
2 & 2 & 2 & 1 &
2 & 2 & 2 & 2 \\
\hline
\hline
\multirow{3}{*}{Exclusivity} &
avg $\mathcal{O}$, avg $\mathcal{C}$ & 
\multicolumn{4}{c||}{(0.12, 0.08)} &
\multicolumn{4}{c||}{(0.24, 0.17)} &
\multicolumn{4}{c||}{(0.13, 0.04)} &
\multicolumn{4}{c|}{(0.09, 0.06)} \\
\cline{3-18}
& \#products &
13 &  6 &  3 & 0 &
 2 &  1 &  1 & 0 &
90 & 29 & 17 & 7 &
35 & 12 &  5 & 0 \\
& \#badges &
1 & 1 & 1 & 0 &
1 & 1 & 1 & 0 &
2 & 2 & 1 & 1 &
1 & 1 & 1 & 0 \\
\hline
\end{tabular}}
\label{tab:longevity_badge}
\end{table*}

\subsection{High consistency for \textit{Scarcity} and \textit{Urgency} badges!}
Along with Occurrence, we also computed Consistency, 
and we were particularly interested in 
badges belonging to the \textit{Scarcity} and \textit{Urgency} categories. Ideally, 
for these two categories, consistency values should be low. In fact, we 
captured multiple daily snapshots 
to track the outflux of these `rare' products. However, the actual scenario is quite different.

For Bewakoof, we found that 9 products (all under `women kurti' or `women jackets') had a \textit{Scarcity} badge for all the snapshots. 
There were 26 products that had a consistency value of more than 0.6 
-- all of these products are for `women'; among products for `men', the highest consistency value was 0.52.
Following a similar pattern, in FirstCry, there were 11 products that had a \textit{Scarcity} badge for all the snapshots, and all of these products appeared under `baby girl dresses'.
For gender-agnostic searches in `Baby Products', products under `stroller' typically had highest consistency values which were around 0.4, barring one (Tiffy \& Toffee Portable Stroller with Canopy) that had a consistency value of 0.69 in FirstCry. 
In Hopscotch, the number of products that had a \textit{Scarcity} badge all throughout is 26 -- however, most of these belonged to `onesies' and `baby footwear'.
Moreover, the consistency values for the products under `baby boy clothes' and `baby girl dresses' were comparable for this website.
Contrarily, the generic websites typically had lower consistency values, especially Amazon, that had a `sofa set' with the highest consistency value of 0.71.
Snapdeal, however, had a handful of products (8) with consistency values above 0.9; two of these were `face primer' and the rest belonged to `sofa set' or `dining set'. 
Although we found that the highest consistency values in the generic websites mostly belonged to `Home Decor' domain, 
surprisingly, the niche website Urban Ladder had lower consistency values with the highest one being 0.61 belonging to a `table lamp'.

In contrast to \textit{Scarcity} badges, \textit{Urgency} badges have relatively lower consistency values.
For the websites Nykaa, LimeRoad and Pepperfry, the average 
consistency values were 0.2, 0.3 and 0.32 respectively. 
For Bewakoof, one product under `men polo t-shirt' had a consistency value of 0.92. 
My Baby Babbles, however, had high consistency values ($> 0.88$) for almost all products under `baby shampoo', `baby towels' and `baby diaper'. 
Three products in Amazon -- all moisturizers from Mamaearth -- had urgency badge in all 90 snapshots. See Table~\ref{tab:archit_table} for more examples.

\begin{table}[t!]
\centering
\tbl{\bf Examples of products with high consistency values.}
{
\begin{tabular}{|l||l||l|}
\hline
{\bf Website} & {\bf Product} & {\bf Badge}\\
\hline
Bewakoof & Women's Sleeveless Button Down Mandarin Collar Side Slit Kurta & Few Left \\
Bewakoof & Jet Black Plain Sleeveless Puffer Jacket with Detachable Hood & Few Left\\
Bewakoof & Women's Rib Raglan Dress & Buy X Get Y \\
FirstCry & Baby Peony Dress Set & X Left \\
FirstCry & TINY BABY Sleeveless Checked Dress & X Left\\
Hopscotch & Pink Flower Lace Onesies with Headband & X Left \\
Hopscotch & Navy Blue LED Shoes & X Left \\
My Baby Babbles & 3 Sprouts Stroller Organiser & Sale \\
My Baby Babbles & My Milestones Baby Hooded Towel - Modern Stripped & Sale \\
Snapdeal & Mars 2in1 Makeup Fixer Spray \& Face Primer Gel & X Left! \\
Snapdeal & Mee Mee Pram Cum Stroller With Rocking Function & X Left! \\
e.l.f. Cosmetics & No Budge Precision Eyeliner & New \\
e.l.f. Cosmetics & Jelly Poppin' Skincare Set & New \\
Max Fashion & MAX Printed Straight Kurta & New \\
Max Fashion & MAX Solid Slim Fit Casual Shirt & New \\
\hline
\end{tabular}}
\label{tab:archit_table}
\end{table}

\section{RQ2. Uniformity: Did you get the same deal?}
Next, we investigate whether products sold on multiple platforms are assigned similar badges. 
We found that if a product is present in multiple platforms, the product names are similar (if not exactly the same) across websites, and contain the brand name and product specifications like color, fiber (for apparels), etc. For example, ``Lakme Insta Eye Liner, Black, Water Resistant, Long-Lasting, 9 ml" on Amazon and ``Lakme Insta Eye Liner - Black" on Nykaa. To account for the possibility of not finding an exact name match, we match the products on different platforms based on parameters like brand, color, etc. Moreover, even the names of the badges are often unique across the platforms, thus, we look for matches in the badge category. 

Surprisingly, out of the 29 possible platform-pairs, we found that only 10 of these had at least one product that was assigned a badge on both, as mentioned in Table~\ref{tab:uniformity_website}. In this table, we mention for each platform-pair, the number of queries (out of 40 for Amazon-Snapdeal pair, and out of 10 for every other pair) for which there is at least one common product that has received the same badge category, the number of different badge categories that these products belong to, and two scores: $Score_P$ and $Score_L$, which we explain with the following example.

Let us consider that there are two platforms $X$ and $Y$.
For simplicity, let there be a single badge category $\mathcal{B}$.
Now let the products having a $\mathcal{B}$-type badge and their corresponding {\it Occurrence} in the two platforms be as follows: 
$X:: \{A:0.3, B:1.0, C:0.4, D:0.2 \}$, and $Y:: \{A:0.2, C:0.6, E:1.0, F:0.1\}$.
We compute $Score_P$ as the ratio of the number of common products to the number of all products in the two platforms belonging to the same badge category. In this example, 

$Score_P = \frac{|\{x: x \in X \land x \in Y \}|}{|\{x: x \in X \lor x \in Y\}|} = \frac{|\{A,C\}|}{|\{A,B,C,D,E,F\}|} = \frac{2}{6}$ = 0.33.

On the other hand, we define $Score_L$ as the ratio of the {\it Occurrence} of common products to the {\it Occurrence} of all products in the two platforms. Here

$Score_L = \frac{\Sigma_{\{x: x \in X \land x \in Y\}} \mathcal{O}_{\mathcal{B},x}}{\Sigma_{\{x: x \in X \lor x \in Y\}} \mathcal{O}_{\mathcal{B},x}} = \frac{0.3+0.2+0.4+0.6}{0.3+1.0+0.4+0.2+0.2+0.6+1.0+0.1} = \frac{1.5}{3.8} = 0.3947$. \\
Note that a higher $Score_P$ does not necessarily imply a higher $Score_L$, or vice versa.

Predictably, the maximum number of queries for which there is some common product with identical badge category occurs for the generic platform-pair: Amazon-Snapdeal.
In terms of $Score_P$ and $Score_L$, we see that the pair Amazon-Nykaa has the maximum uniformity.
If we consider only those platform-pairs where both websites are niche, then we find Nykaa-e.l.f. Cosmetics to have the maximum uniformity in terms of both the scores.
These observations suggest that there is highest uniformity in the \textit{Cosmetics} domain compared to other domains considered here.

Table~\ref{tab:uniformity_badge} shows the summary of product badges based on badge categories.
For \textit{Recency} and \textit{Exclusivity} badges, there was no overlap between any platform-pairs, and hence these categories do not appear in Table~\ref{tab:uniformity_badge}.
Based on $Score_P$ and $Score_L$, \textit{Social Proof} badges are the most common -- this result may seem intuitive because popular brands should be in demand irrespective of platforms, and the longevity of \textit{Social Proof} badges is normally higher as can be seen from Table~\ref{tab:longevity_badge}.
On the other hand, \textit{Scarcity} badges seem to be assigned most uniformly across platform-pairs and queries.

\begin{table}[t!]
\centering
\tbl{\bf Uniformity of badges across platform-pairs.}
{
\begin{tabular}{|l|r|r|r|r|}
\hline
{\bf Platform-Pair} & {\bf No. of queries} & {\bf No. of badge categories} & $\mathbf{Score_P}$ & $\mathbf{Score_L}$\\
\hline
Amazon-FirstCry        & 5 & 2 & 0.0399 & 0.0884\\
Amazon-Hopscotch       & 2 & 1 & 0.0371 & 0.0518\\
Amazon-My Baby Babbles & 1 & 1 & 0.0233 & 0.0798\\
{\bf Amazon-Nykaa}     & 9 & 3 & {\bf 0.0932} & {\bf 0.2509}\\
Snapdeal-FirstCry      & 3 & 1 & 0.0110 & 0.0227\\
Snapdeal-Hopscotch     & 5 & 1 & 0.0157 & 0.0213\\
Amazon-Snapdeal        &20 & 3 & 0.0162 & 0.0402\\
FirstCry-Hopscotch     & 2 & 1 & 0.0032 & 0.0078\\
Nykaa-e.l.f. Cosmetics & 1 & 1 & 0.0333 & 0.1313\\
LimeRoad-Bewakoof      & 1 & 1 & 0.0082 & 0.0066\\
\hline
\end{tabular}}
\label{tab:uniformity_website}
\end{table}

\begin{table}[t!]
\centering
{
\caption{\bf Uniformity of badges from different categories.}
\resizebox{.95\columnwidth}{!}
{
\begin{tabular}{|l|r|r|r|r|}
\hline
{\bf Category} & {\bf No. of platform-pairs} & {\bf No. of queries} & $\mathbf{Score_P}$ & $\mathbf{Score_L}$\\
\hline
{\bf Social Proof} & 3 & 13 & {\bf 0.1113} & {\bf 0.2147}\\
{\bf Scarcity}     & {\bf 6} & {\bf 22} & 0.0173 & 0.0341\\
Urgency      & 2 & 10 & 0.0592 & 0.2044\\
Promotional  & 1 &  1 & 0.0082 & 0.0066\\
Endorsement  & 3 &  7 & 0.0505 & 0.2255\\
\hline
\end{tabular}}
\label{tab:uniformity_badge}}
\end{table}

\section{RQ3. Assignment bias: What is so special about the product?}
Since having a badge helps a product get the desired traction with the customers, it is pertinent to ask whether 
the products with badges are any different from those which did not get badges. 
This can also help in predicting the underlying badge assignment policies employed by different websites.
We look into the following four attributes: 
price, discount, average customer rating (CR), and number of customer ratings to compare the products.
To clarify the attribute `discount', 
if the original price of a product is $X$ but currently it is available at a lower price $Y$, then the discount 
is $\frac{X-Y}{X}$. 
It is important to note that 
e-commerce websites can give discounts without any explicit \textit{Promotional} badges. 
Moreover, in such cases, we consider the `price' of the product to be its current selling price, i.e., $Y$ in our example.

\begin{figure*}[t]
 \centering
  \subfloat[\bf Scheme 1]{
  \centering
  \includegraphics[width=\textwidth]{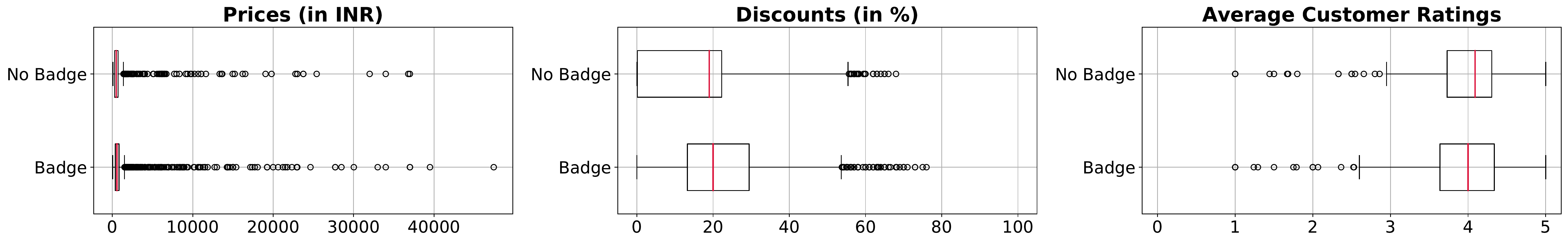}
  \label{fig:scheme1}}
 \hfill
  \subfloat[\bf Scheme 2]{
  \centering
  \includegraphics[width=\textwidth]{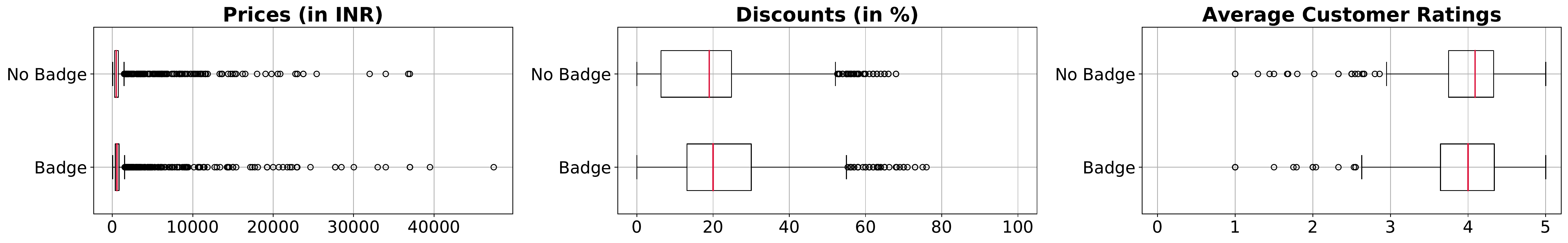}
  \label{fig:scheme2}}
 \caption{\bf Comparing products with and without badges in FirstCry website, following two different schemes (details in text).}
 \label{fig:scheme}
\vspace*{+4mm}
\end{figure*}

Since the badge assignment to a product can vary across snapshots, we followed two schemes while creating the list of products with `Badge' and with `No Badge': 1. we include a product in the `Badge' list if it has received a badge in at least one snapshot; 2. we include a product in the `Badge' list only for those snapshots where it has received a badge, and do not include 
for snapshots where it did not get a badge. 
Figure~\ref{fig:scheme} shows the comparison between these two types of products in both schemes, for the FirstCry website. Interestingly, there is no significant difference in the comparison results %
between these two schemes. For both, products with badges offer higher discounts; a few of them are higher-priced, although the median price is similar for both types. Surprisingly, we found that the average rating is higher for products without badges, raising concerns about whether lower quality products are being promoted using the badges. We found similar trend for other platforms as well.


\begin{table*}[t!]
\centering
\caption{\bf Summary of differences (measured using KS statistic) between products with and without a particular badge category. In each cell, the most distinguishing attribute is highlighted.}
\vspace{1mm}
\resizebox{0.98\linewidth}{!}
{
\begin{tabular}{|l|r|| r|r|r|r|r|r|r|}
\hline
& Category & Social Proof & Scarcity & Urgency & Recency & Promotional & Endorsement & Exclusivity \\
\hline
\multirow{4}{*}{Amazon} &
Price & 
0.1258 & {\bf 0.0980} & 0.0490 & & & 0.2425 & {\bf 0.4123} \\
& Discount &
0.0870 & 0.0654 & 0.1158 & & & 0.1037 & 0.2798 \\
& Avg. CR &
0.1758 & 0.0308 & 0.1444 & & & 0.4098 & 0.2312 \\
& \#CR &
{\bf 0.6352} & 0.0465 & {\bf 0.4522} & & & {\bf 0.6881} & 0.3722 \\
\hline
\multirow{2}{*}{Bewakoof} &
Price & 
& {\bf 0.1002} & {\bf 0.7709} & & 0.1077 & & \\
& Discount &
& 0.0796 & 0.7638 & & {\bf 0.1248} & & \\
\hline
\multirow{4}{*}{e.l.f. Cosmetics} &
Price & 
0.2004 & & & 0.2427 & & 0.5661 & \\
& Discount &
0.1475 & & & 0.1951 & & 0.1455 & \\
& Avg. CR &
0.2663 & & & 0.2288 & & 0.3500 & \\
& \#CR &
{\bf 0.7143} & & & {\bf 0.5148} & & {\bf 0.5864} & \\
\hline
\multirow{4}{*}{FirstCry} &
Price & 
0.2349 & 0.1738 & & 0.2389 & 0.3443 & & \\
& Discount &
0.2051 & {\bf 0.2107} & & 0.1715 & 0.1970 & & \\
& Avg. CR &
0.0993 & 0.1435 & & {\bf 0.2480} & {\bf 0.6307} & & \\
& \#CR &
{\bf 0.7222} & 0.0836 & & 0.1398 & 0.0919 & & \\
\hline
\multirow{2}{*}{Hopscotch} &
Price & 
& {\bf 0.1941} & & & & & \\
& Discount &
& 0.1800 & & & & & \\
\hline
\multirow{2}{*}{LimeRoad} &
Price & 
& & 0.0632 & {\bf 0.1440} & 0.1118 & & {\bf 0.2490} \\
& Discount &
& & {\bf 0.1364} & 0.1264 & {\bf 0.1455} & & 0.1856 \\
\hline
\multirow{2}{*}{Max Fashion} &
Price & 
& & & 0.0795 & 0.1458 & & \\
& Discount &
& & & {\bf 0.2625} & {\bf 0.4699} & & \\
\hline
\multirow{2}{*}{My Baby Babbles} &
Price & 
& & 0.2182 & & & & \\
& Discount &
& & {\bf 1.0000} & & & & \\
\hline
\multirow{4}{*}{Nykaa} &
Price & 
0.2192 & & 0.0920 & 0.3052 & 0.0420 & 0.0663 & \\
& Discount &
0.2034 & & 0.1305 & 0.2755 & 0.1606 & 0.0794 & \\
& Avg. CR &
0.2272 & & 0.1509 & 0.3266 & 0.1249 & 0.0774 & \\
& \#CR &
{\bf 0.5721} & & {\bf 0.2483} & {\bf 0.3469} & {\bf 0.2152} & {\bf 0.1470} & \\
\hline
\multirow{2}{*}{Pepperfry} &
Price & 
{\bf 0.1581} & & {\bf 0.5005} & {\bf 0.4067} & {\bf 0.3366} & & \\
& Discount &
0.1324 & & 0.2163 & 0.2229 & 0.3255 & & \\
\hline
\multirow{4}{*}{Snapdeal} &
Price & 
0.2218 & 0.1609 & & & & 0.0597 & \\
& Discount &
0.4038 & 0.0720 & & & & 0.0582 & \\
& Avg. CR &
0.2771 & {\bf 0.1642} & & & & {\bf 0.5389} & \\
& \#CR &
{\bf 0.7142} & 0.0730 & & & & 0.3795 & \\
\hline
\multirow{2}{*}{Urban Ladder} &
Price & 
{\bf 0.2926} & {\bf 0.2312} & & 0.3380 & & & \\
& Discount &
0.2071 & 0.2284 & & {\bf 0.4585} & & & \\
\hline
\end{tabular}
}
\label{tab:KStest}
\end{table*}

Next, 
for a given platform, we compare the products with badges belonging to a particular badge category against those without a similar badge (even though some of these products may have badges but of a different category). 
We used the Kolmogorov–Smirnov (KS) statistic~\citep{lilliefors1967kolmogorov} to measure the difference between the distributions of price, rating etc.; a summary of our findings can be found in Table~\ref{tab:KStest}.
In this table, majority of the cells are empty because none of the websites we investigated provided badges across all categories; in fact, the maximum badge categories covered by an individual platform is 5 (Amazon, Nykaa), and the minimum is 1 (Hopscotch, My Baby Babbles).
Moreover, only 3 of the niche websites (e.l.f. Cosmetics, FirstCry, Nykaa) out of 10, listed the average customer ratings and the number of ratings.
In each cell in Table~\ref{tab:KStest}, we highlight the highest KS statistic that signifies which attribute is most helpful to distinguish products having badges of a specific category with others.

\begin{figure}[t]
\centering
\includegraphics[width=0.5\columnwidth]{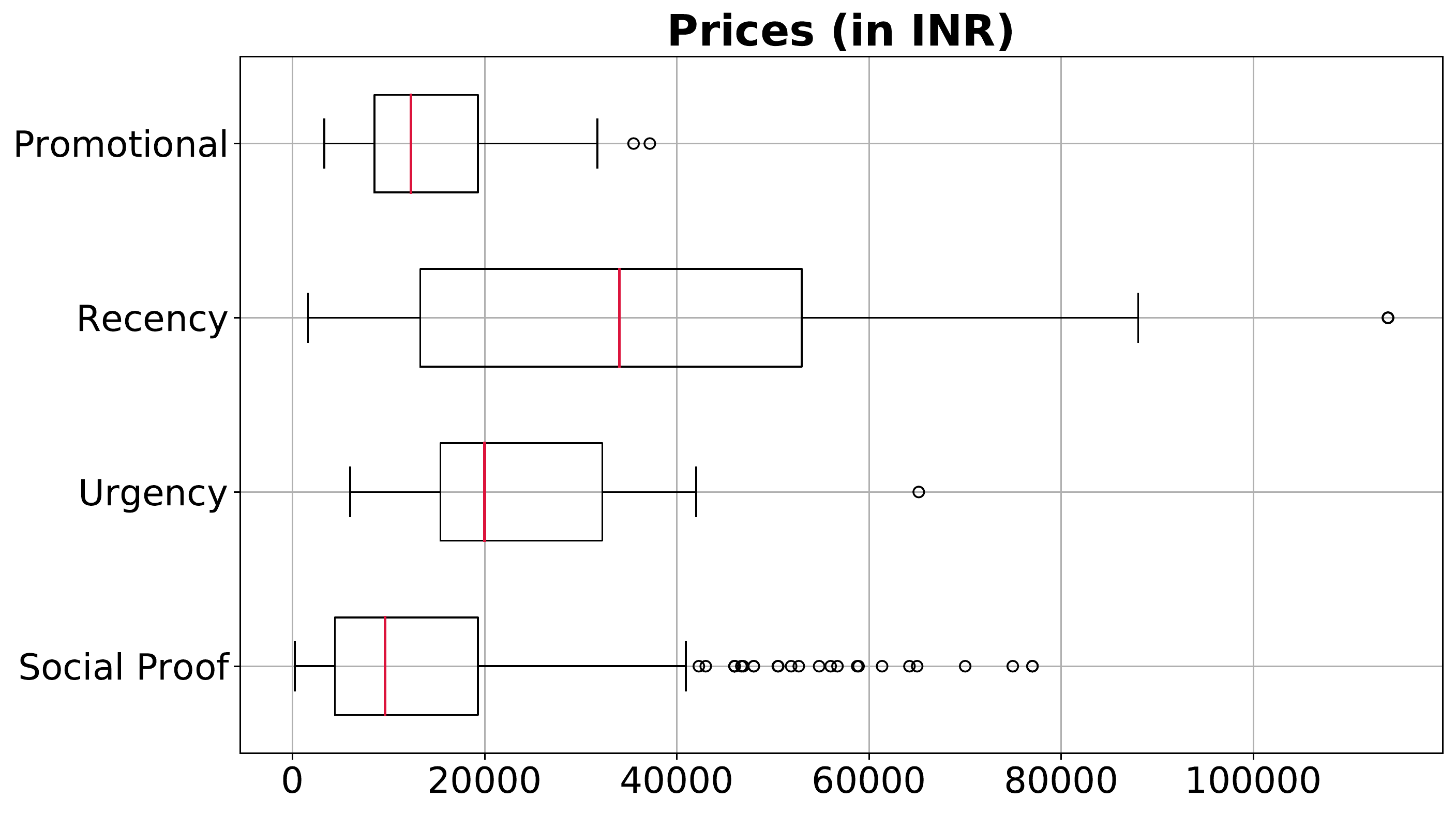}
\caption{\bf Box plots for Pepperfry: X-axis denotes  prices, and Y-axis denotes badge categories.}
\label{fig:pepperfry_price}
\end{figure}

We find that the number of customer ratings is the most distinguishing attribute for \textit{Social Proof} badges because the products which have these badges usually have more ratings, thus implying more popularity.
Similarly, since recently added products have much lesser customer ratings, even for \textit{Recency} badges, the number of customer ratings is the most distinguishing attribute, whenever available, with the exception of FirstCry website.
Oddly, \textit{Endorsement} badges are also most distinguishable based on the count of customer ratings for most websites.
For the only two websites that give away \textit{Exclusivity} badges, price is the most distinguishing attribute.
There does not seem to be any specific trend related to the most distinguishing attribute when it comes to \textit{Scarcity}, \textit{Urgency} and \textit{Promotional} badges.
However, it may be pertinent to note that My Baby Babbles has a perfect KS statistic of 1.0 with respect to the discount attribute because this site gives away only one type of badge `Sale' (\textit{Urgency}), and assigns it to any product that is giving discount.
The number of customer ratings happens to be the most distinguishing attribute across all badge categories for Nykaa. A similar issue occurs with Pepperfry where price is the most distinguishing attribute -- 
if we draw box plots for the badge categories based on price, these should be separable as shown in Figure~\ref{fig:pepperfry_price}. 

\begin{table}[t]
\centering
\caption{\bf Number of instances where multiple badges were assigned to the same product in a platform.}{
\begin{tabular}{|l|c|}
\hline
{\bf Platform}  & {\bf No. of instances}\\
\hline
Amazon       & 2-679, 3-74, 4-8\\
Bewakoof     & 2-192, 3-16\\
FirstCry     & 2-154, 3-3\\
LimeRoad     & 2-752, 3-53\\
Max Fashion  & 2-207, 3-45\\
Nykaa        & 2-773, 3-190, 4-23\\
Pepperfry    & 2-42\\
Snapdeal     & 2-685, 3-65\\
Urban Ladder & 2-214\\
\hline
\end{tabular}}
\label{tab:multi_platform}
\end{table}

\subsection{Products with multiple badges} 

In addition to the earlier experiments, we dived into the recommendation lists to see which products received multiple badges in a platform.
We found that 9 out of 12 websites assigned 2, 3 or 4 badges to a single product (in different snapshots or, in some rare cases, even in the same snapshot).
For example, in Amazon, Harpa Women's A-Line Dress received 4 badges: Amazon's Choice (\textit{Endorsement}), Best Seller (\textit{Social Proof}), Only X Left in Stock (\textit{Scarcity}) and Deal of the Day (\textit{Urgency}).
Similarly, in Nykaa, Lakme Eyeconic Liquid Eyeliner received 4 badges: Featured (\textit{Endorsement}), Bestseller (\textit{Social Proof}), Sale (\textit{Urgency}) and Offer (\textit{Promotional}).
These results are captured in Table~\ref{tab:multi_platform}.

In Table~\ref{tab:multi_badge}, we provide the number of instances, accumulated over all platforms, where two badge categories were assigned to the same product. Thus, if a product receives badges of three different categories, then we will add it to the count of all the three possible category-pairs. Note that no product was assigned both \textit{Endorsement} and \textit{Exclusivity} badges, and hence only this pair is missing in Table~\ref{tab:multi_badge}. From this table, it seems that websites take a two-pronged approach -- \textit{Urgency} and \textit{Promotional} -- to quickly sell specific products.
Furthermore, we conjecture that popular products having \textit{Social Proof} badges may often sell fast, and when the resources are depleted, \textit{Scarcity} badges are assigned to these. We think \textit{Recency} and \textit{Promotional} badges probably go hand-in-hand to mitigate the cold start problem in recommendations. Similar opinions may be formed for other category-pairs with high counts. 

\begin{table}[t!]
\centering
\tbl{\bf Number of instances where a badges category pair was assigned to the same product.}
{
\begin{tabular}{|l|r||l|r|}
\hline
{\bf Category-Pair}  & {\bf No. of instances} & {\bf Category-Pair} & {\bf No. of instances}\\
\hline
Urgency-Promotional      & 1074 & Social Proof-Scarcity    & 748\\
Recency-Promotional      &  485 & Scarcity-Urgency         & 480\\
Urgency-Recency          &  353 & Social Proof-Urgency     & 276\\
Scarcity-Recency         &  261 & Urgency-Endorsement      & 253\\
Scarcity-Endorsement     &  214 & Social Proof-Endorsement & 178\\
Promotional-Endorsement  &  173 & Scarcity-Promotional     & 121\\
Social Proof-Promotional &   56 & Promotional-Exclusivity  &  54\\
Urgency-Exclusivity      &   26 & Social Proof-Recency     &  10\\
Scarcity-Exclusivity     &   10 & Recency-Endorsement      &   5\\
Social Proof-Exclusivity &    2 & Recency-Exclusivity      &   2\\
\hline
\end{tabular}}
\label{tab:multi_badge}
\end{table}

\section{Conclusion}
\label{sec:concl}
Assigning badges to products is an effective way to highlight specific products in any e-commerce website.
Depending on the badge assigned, a customer may experience different emotions on encountering it, such as assurance on seeing `Best Seller' or panic on observing `Few Left'.
These emotions may nudge her to buy the product with a particular badge instead of others. In spite of 
the effect on both the customers and the sellers of the e-commerce platforms, there was no systematic cross-platform study on product badges. In this work, we took the first attempt to bridge this gap.
We first categorized different product badges based on the emotions they trigger.
Then, we examined the badges assigned by different generic and niche e-commerce platforms, catering to four diverse domains. We particularly focused on the longevity of badge assignments, their uniformity across platforms, and how these vary based on platforms, badge categories and domains.
In the process, we try to identify the trends, and point out interesting idiosyncrasies in some cases.

This work has several takeaways 
for the different stakeholders in an e-commerce business, namely, the customers, the sellers and the marketplace platform provider. 
For the customers, firstly, they should be aware of the different emotional reactions that the badges try to trigger -- being aware should save them from falling an easy prey; for example, some items may not be as scarce as the \textit{Scarcity} badges would like one to believe.
Secondly, products with badges are not necessarily better in quality than those without it, and hence, they should ideally buy after careful comparison.

For the sellers, although they may vie for product badges to get that extra boost in sales, they should bear in mind that badge assignment, especially when it comes to the \textit{Endorsement} category, does not necessarily reflect that the products with badges are superior to those without. Moreover, securing a badge in a platform is also unlikely to guarantee a similar success in another platform.

For the e-commerce platforms, first and foremost, they should make the badge assignment process transparent  to achieve a level playing field for all the sellers, and thus prevent dissatisfaction among smaller sellers.
Furthermore, they should closely regulate badge assignments to restrain some products from retaining a badge for a prolonged period, especially for \textit{Scarcity}, \textit{Urgency} and \textit{Recency} badges.
If a standardization of product badge assignment policies can be achieved, it would allow more uniformity across platforms as well, which would provide extra impetus to the sellers to ensure that their products are of high quality, and this, in turn, would benefit the customers.

In this work, we didn't consider the personalization of product badges -- whether they exist and if yes, to what extent. In future work, we want to check if badge assignments differ based on the target demography's gender, age, etc. We would also like to carry out a more extensive study including e-commerce platforms from other countries, adding more queries and snapshots. We hope that in the long run, this line of work will help achieve more accountability in product badge assignments across e-commerce platforms, helping customers and small sellers alike. 

\bibliographystyle{tfcad}
\bibliography{main}

\end{document}